\documentclass[twocolumn,conference]{IEEEtran}
\usepackage[LGR,T1]{fontenc}
\usepackage[latin9]{inputenc}
\usepackage{float}
\usepackage{amsmath}
\usepackage{amssymb}
\usepackage{stmaryrd}
\usepackage{graphicx}
\usepackage{wasysym}
\usepackage{esint}
\usepackage[unicode=true,
 bookmarks=true,bookmarksnumbered=true,bookmarksopen=true,bookmarksopenlevel=1,
 breaklinks=false,pdfborder={0 0 0},pdfborderstyle={},backref=false,colorlinks=false]
 {hyperref}
\hypersetup{pdftitle={Your Title},
 pdfauthor={Your Name},
 pdfpagelayout=OneColumn, pdfnewwindow=true, pdfstartview=XYZ, plainpages=false}

\makeatletter

\DeclareRobustCommand{\greektext}{%
  \fontencoding{LGR}\selectfont\def\encodingdefault{LGR}}
\DeclareRobustCommand{\textgreek}[1]{\leavevmode{\greektext #1}}
\ProvideTextCommand{\~}{LGR}[1]{\char126#1}

\floatstyle{ruled}
\newfloat{algorithm}{tbp}{loa}
\providecommand{\algorithmname}{Algorithm}
\floatname{algorithm}{\protect\algorithmname}

\usepackage[caption=false,font=footnotesize]{subfig}

\makeatother

\begin{document}
\title{Multi-User Localization and Tracking with Spatiotemporal Correlation
in Multi-RIS-Assisted Systems}
\author{\IEEEauthorblockN{Ronghua Peng$^{\dagger}$, Peng Gao$^{\dagger\ddagger*}$,
Jing You$^{\dagger}$, Lixiang Lian$^{\dagger}$}\IEEEauthorblockA{$^{\dagger}$School
of Information Science and Technology, ShanghaiTech University, Shanghai
201210, China\\$^{\ddagger}$Innovation Academy for Microsatellites
of CAS, Shanghai 201210, China\\$^{*}$University of Chinese Academy
of Sciences, Beijing 100049, China\\Email: \{pengrh, gaopeng, youjing2023,
lianlx\}@shanghaitech.edu.cn}\thanks{Corresponding author: Lixiang Lian.}}
\maketitle
\begin{abstract}
As a promising technique, reconfigurable intelligent surfaces (RISs)
exhibit its tremendous potential for high accuracy positioning. In
this paper, we investigates multi-user localization and tracking problem
in multi-RISs-assisted system. In particular, we incorporate statistical
spatiotemporal correlation of multi-user locations and develop a general
spatiotemporal Markov random field model (ST-+MRF) to capture multi-user
dynamic motion states. To achieve superior performance, a novel multi-user
tracking algorithm is proposed based on Bayesian inference to effectively
utilize the correlation among users. Besides that, considering the
necessity of RISs configuration for tracking performance, we further
propose a predictive RISs beamforming optimization scheme via semidefinite
relaxation (SDR). Compared to other pioneering work, finally, we confirm
that the proposed strategy by alternating tracking algorithm and RISs
optimization, can achieve significant performance gains over benchmark
schemes. 
\end{abstract}

\IEEEpeerreviewmaketitle{}

\section{Introduction}

As one of the key technologies of sixth-generation (6G), integrated
sensing and communication (ISAC) can effectively address the issue
of spectrum scarcity, reduce hardware cost and further promote the
network intelligence in 6G system \cite{ISAC9737357}. Well-shaped
communication signals can be employed to perform various sensing tasks.
In ISAC-enabled millimeter-wave (mmWave) massive MIMO systems, highly
directional communication waveforms supports high-precision localization
services\cite{Device-Free9724258}. However, traditional wireless
localization algorithms heavily rely on the presence of a line-of-sight
(LoS) path to extract useful location information \cite{Gao9898900}.
Reconfigurable intelligent surfaces (RISs) can be effectively utilized
to aid localization by altering the propagation direction of signals
in obstructed LoS scenarios. Multiple RISs with known positions can
also serve as virtual anchors to support direct localization \cite{Multiple-RISs10041765},
which achieves superior performance compared to the conventional indirect
localization techniques, such as angle-assisted triangulation or distance-assisted
trilateration \cite{He9348112}.

Single RIS-assisted localization has been studied in \cite{He9348112,Zhong10042240,JointT9769997,yu2023active,9511765}
, where indirect localization methods were studied in \cite{He9348112,Zhong10042240,JointT9769997}
and direct location techniques were investigated in \cite{yu2023active,9511765}
. Compared to single RIS systems, multiple RISs-aided localization
can effectively improve the coverage and reliability of localization
services and has been studied in \cite{9511765}. In particular, the
lower bound of position estimation error was analyzed in \cite{Multiple-RISs10041765}
in a multi-RIS single-user localization system to optimize the RIS
design. Similarly, position and rotation error bounds were explored
in \cite{CRB9977919} in multi-RIS and multi-user scenarios, which
were utilized to optimize the beamforming (BF) of base station (BS)
and RISs. However, these papers have not investigated how to effectively
locate multiple users in a multi-RIS system, especially when users
are moving. Single user location tracking has been considered in \cite{Bayesian9772371},
where the temporal correlation of user's location has been exploited
to design a Bayesian-based user localization and tracking algorithm
(BULT) in multi-RIS systems. The Bayesian Cramer Rao bound (BCRB)
was adopted to characterize the fundamental performance limitations
of the considered tracking problem. Nonetheless, the extension to
the multi-user scenario is challenging. Firstly, when users' locations
are not only temporal but also spatially correlated, such as a group
of users or vehicles are dynamically moving, it is challenging to
construct a model to capture such spatiotemporal correlations and
design a dynamic positioning algorithm to realize multi-user direct
localization and tracking in multi-RIS systems. Secondly, analyzing
the positioning error in multi-user multi-RIS systems in the presence
of spatiotemporal correlation and optimizing the passive BF of RISs
pose additional challenges.

In this paper, we consider multi-user localization and tracking problem
in a multi-RIS-assisted mmWave system with spatiotemporal correlations.
Firstly, we employ Markov Random Field (MRF) to model the motion trajectories
of multiple users to leverage the potential correlations in both temporal
and spatial domain within the multi-user positions. We show that such
model is general enough to encompass various practical applications.
Secondly, we propose a multi-user direct localization and tracking
(MUDLT) algorithm based on the Bayesian inference to effectively utilize
the spatiotemporal correlations among users' locations. Last but not
the least, we analyze the spatiotemporal correlation assisted positioning
error bound (ST-PEB) in dynamic multi-user localization system, which
is leveraged to optimize the passive BF at multi-RISs to provide optimal
positioning services. Matrix lifting and semidefinite relaxation (SDR)
techniques are adopted to solve the resulting nonconvex optimization
problem. Finally, we demonstrate the superior performance of proposed
MUDLT algorithm and passive BF design through numerical experiments.

\section{System Model }

\begin{figure}[tbh]
\begin{centering}
\includegraphics[scale=0.24]{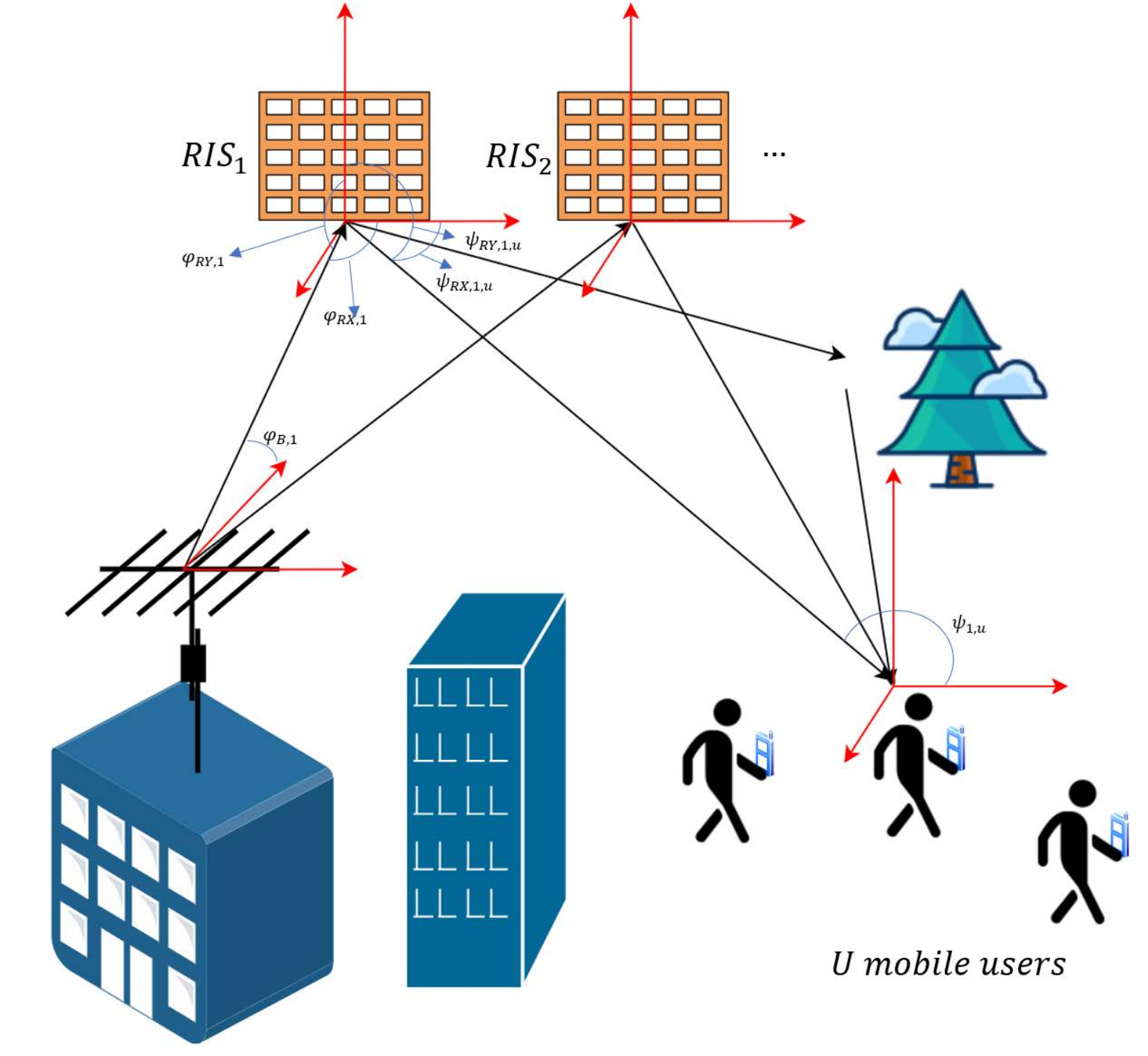}
\par\end{centering}
\caption{System model of multi-RIS-assisted multi-user localization and tracking.\label{Fig. 1}}
\end{figure}

We consider a multi-RIS-assisted mmWave MIMO system, which consists
of one BS with $N_{B}$ antennas, $N$ RISs each with $M_{R}=N_{x}\times N_{y}$
elements, and $U$ users each with $N_{U}$ antennas, as illustrated
in Fig. 1. The antenna/element interval is half of the carrier frequency
wavelength $\lambda$. The virtual-line-of-sight (VLoS) path reflected
by each RIS always exists while the LoS path from BS to user is blocked
by obstacles. We assume users are dynamically moving. The positions
of the BS, the $n$-th RIS and the $u$-th user at the $t$-th time
slot are denoted by the two-dimension vectors $\boldsymbol{U}_{B}$,
$\boldsymbol{U}_{n}$ and $\boldsymbol{U}_{u}^{t}$, respectively.
Consider that the BS and RISs are static in general, and assume $\boldsymbol{U}_{B}$
and $\boldsymbol{U}_{n}$ are known at the BS in advance.

\subsection{Signal Model}

Let $\mathbf{G}_{n}$ and $\mathbf{G}_{n,u}^{t}$ denote the downlink
channel between BS and the $n$-th RIS and the channel between the
$n$-th RIS to the $u$-th user at time slot $t$, respectively. Due
to the significant multi-path degradation in mmWave systems, we assume
LoS path between BS, RIS and user.Therefore, $\mathbf{G}_{n}$ and
$\mathbf{G}_{n,u}^{t}$ are given by

\begin{equation}
\mathbf{G}_{n}=\xi_{n}\mathbf{a}_{R}(\varphi_{RX,n},\varphi_{RY,n})\mathbf{a}_{B}^{H}(\varphi_{B,n}),\label{eq:ch1}
\end{equation}

\begin{equation}
\mathbf{G}_{n,u}^{t}=\text{\textgreek{\char6}}_{n,u}^{t}\mathbf{a}_{U}(\psi_{n,u}^{t})\mathbf{a}_{R}^{H}(\psi_{RX,n,u}^{t},\psi_{RY,n,u}^{t})^{H},\label{eq:ch2}
\end{equation}
where $\xi_{n}$ and $\text{\textgreek{\char6}}_{n,u}^{t}$ represent
the corresponding channel gains, $\boldsymbol{a}_{R}(\cdot)\in\mathbb{C}^{M_{R}\times1}$,
$\boldsymbol{a}_{B}(\cdot)\in\mathbb{C}^{N_{B}\times1}$ and $\boldsymbol{a}_{U}(\cdot)\in\mathbb{C}^{N_{U}\times1}$
represent the steering vectors of the RIS, BS and user, respectively.
And $\boldsymbol{a}_{x}(\theta)=[1,e^{j\pi\theta},\ldots,e^{jN_{x}\theta}]$
with $x\in\{R,B,U\}$, $N_{x}$is corresponding number of antennas.
$\varphi_{RX,n}$ and $\psi_{RX,n}^{t}$ are the cosine of the angle-of-arrivals
(AoA) and angle of departure (AoD) in horizontal direction for the
$n$-th RIS, while $\varphi_{RY,n}$ and $\psi_{RY,n}^{t}$ are AoA
and AoD of the $n$-th RIS in azimuth direction, respectively. $\varphi_{B,n}$
and $\psi_{n,u}^{t}$ represent the AoD of the BS and the AoA of the
$u$-th user at time $t$. Notice that $\varphi_{RX,n}$, $\varphi_{RY,n}$,$\varphi_{B,n}$
are known since the positions of RISs and BS are known. The angles
mentioned above can be calculated by geometric relationship as in
\cite{Bayesian9772371}. For example, $\psi_{n,u}^{t}$ can be calculated
by
\begin{equation}
\psi_{n,u}^{t}=\frac{(\boldsymbol{U}_{u}^{t}-\boldsymbol{U}_{n})^{T}\boldsymbol{e}_{U}}{\bigparallel\boldsymbol{U}_{u}^{t}-\boldsymbol{U}_{n}\bigparallel_{2}},
\end{equation}
where $\boldsymbol{e}_{U}$ is the unit direction vector of the user
antennas and obtained by the built-in sensor in advance. Based on
the channel model in \eqref{eq:ch1} and \eqref{eq:ch2}, the received
signal at user $u$ at time $t$ can be expressed as:

\begin{equation}
\mathbf{y}_{u}^{t}=\sum_{n=1}^{N}\rho_{n}^{t}\mathbf{G}_{n,u}^{t}\mathbf{\Phi}_{n}^{t}\mathbf{G}_{n}\mathbf{w}^{t}x^{t}+\mathbf{n}_{u}^{t},\label{eq:channel_model}
\end{equation}
where $\mathbf{\Phi}_{n}^{t}=\mathrm{diag}(\mathbf{\mathbf{\boldsymbol{\lambda}}}_{n}^{t})\in\mathbb{C}^{M_{R}\times M_{R}}$
is the diagonal phase shift matrix of the $n$-th RIS with $\mathbf{\mathbf{\boldsymbol{\lambda}}}_{n}^{t}=[\lambda_{n,1}^{t},...,\lambda_{n,M_{R}}^{t}]^{T}$
and $|\lambda_{n,i}^{t}|=1,\forall i,n,t$; $\mathbf{w}^{t}\in\mathbb{C}^{N_{b}}$
is the BF vector at BS side; $x^{t}$ is the pilot signal transmitted
by BS and is set as $1$ without loss of generality; $\mathbf{n}_{u}^{t}\sim CN(\mathbf{n}_{u}^{t};0,\sigma_{n}^{2}\mathbf{I})$
is the added Gaussian noise at user antennas, and $\rho_{n}^{t}$
is the reflection coefficient at the $n$-th RIS  in different AoAs
and AoDs\cite{tang2020wireless}.. Then \eqref{eq:channel_model}
can be further simplified as 
\begin{equation}
\mathbf{y}_{u}^{t}=\sum_{n=1}^{N}\bar{\rho}_{n,u}^{t}\mathbf{a}_{U}(\psi_{n,u}^{t})+\mathbf{n}(t),\label{eq:user_signal}
\end{equation}
with

\begin{equation}
\begin{array}{cc}
\bar{\rho}_{n,u}^{t}=\xi_{n}\text{\textgreek{\char6}}_{n,u}^{t}\mathbf{a}_{R}^{H}(\psi_{RX,n,u}^{t},\psi_{RY,n,u}^{t})\times\\
\mathbf{\Phi}_{n}^{t}\mathbf{a}_{R}(\varphi_{RX,n},\varphi_{RY,n})\mathbf{a}_{B}^{H}(\varphi_{B,n})\mathbf{w}^{t}.
\end{array}
\end{equation}
Based on \eqref{eq:user_signal}, we have converted the received signal
into a brief formula which separates the position related angle $\psi_{n,u}^{t}$
and other nuisance channel-related parameters in a unified notation
$\bar{\rho}_{n,u}^{t}$. We aim at estimating $\{\boldsymbol{U}_{u}^{t}\}$
from $\{\mathbf{y}_{u}^{t}\}$ for $u=1,\cdots U$ and $t=1,\cdots,T$
jointly by exploiting the spatiotemporal correlations within $\{\boldsymbol{U}_{u}^{t}\}$.

\subsection{Statistical Spatiotemporal Correlation Model}

In various sensing applications, multi-user localization and tracking
performance can be improved by exploiting the spatial correlation
among users and temporal correlation of each user. To effectively
capture the spatiotemporal correlations, we adopt the MRF \cite{MRF}
model to characterize the location prior. An MRF is a factored probability
function specified by an undirected graph $(V,E),$ where $V$ stands
for the variable nodes and $E$ specifies the correlation network.
If $(u,j)\in E$, which means there is direct connection between node
$u$ and node $j$, and node $u,j$ are correlated with each other.
The joint probability over the random variables can be factorized
as the product of local potential functions $\phi$ at each node and
interaction potential $\psi$ defined on neighborhood cliques. A widely
adopted MRF is a pairwise MRF, where the cliques are restricted to
pairs of nodes. We adopt the pairwise MRF to model the interactions
between nearby users and successive time slots. Specifically, the
following spatiotemporal location prior based on MRF (ST-MRF) is adopted:

\begin{align}
p(\boldsymbol{U}) & =p(\boldsymbol{U}^{1})\prod_{t=1}^{T}p(\boldsymbol{U}^{t}|\boldsymbol{U}^{t-1})\nonumber \\
 & =\prod_{u=1}^{U}\phi(\boldsymbol{U}_{u}^{1})\prod_{(u,j)\in E}\psi(\boldsymbol{U}_{u}^{1},\boldsymbol{U}_{j}^{1})\label{eq:MRF}\\
 & \times\prod_{t=1}^{T}\prod_{u=1}^{U}p(\boldsymbol{U}_{u}^{t}|\boldsymbol{U}_{u}^{t-1})\prod_{(u,j)\in E}\psi(\boldsymbol{U}_{u}^{t},\boldsymbol{U}_{j}^{t}),\nonumber 
\end{align}
where $\boldsymbol{U}$ and $\boldsymbol{U}^{t}$ denote the collection
of $U$ users' locations across $T$ time slots and at the $t$-th
time slot, respectively. In \eqref{eq:MRF}, the temporal correlation
is captured by the transition probability $p(\boldsymbol{U}_{u}^{t}\mid\boldsymbol{U}_{u}^{t-1})$,
which is designed to follow Gaussian distribution \cite{Tem-Gau-MRF,Tem-Gau-MRF2,Tem-Gau-MRF3},
i.e.,
\begin{equation}
p(\boldsymbol{U}_{u}^{t}|\boldsymbol{U}_{u}^{t-1})=\mathcal{N}(\boldsymbol{U}_{u}^{t};\boldsymbol{U}_{u}^{t-1},\boldsymbol{C}),\label{eq:time_pass}
\end{equation}
where covariance matrix $\boldsymbol{C}$ is assumed to be a diagonal
matrix. The local potential functions \textbf{$\phi(\boldsymbol{U}_{u}^{1})$}
at the first time slot are usually set to one for each user $k$ or
uniform distribution over the whole searching area if the prior information
of initial position is unknown\cite{Tem-first-MRF1,Tem-first-MRF2}.
The spatial correlations are modeled by pairwise potential functions
$\psi(\boldsymbol{U}_{u}^{t},\boldsymbol{U}_{j}^{t}$), which are
designed as the function of inter-user distances \cite{Toy1-square,Toy2-square,Toy1-norm,Toy2-norm},
i.e., $\psi(\boldsymbol{U}_{u}^{t},\boldsymbol{U}_{j}^{t})=f(d_{u,j}^{t})$
with $d_{uj}^{t}=\left\Vert \boldsymbol{U}_{u}^{t}-\boldsymbol{U}_{j}^{t}\right\Vert $.
In the following, we will give two toy examples of probability model
\eqref{eq:MRF} in different applications.

\subsubsection{Pedestrian Surveillance}

For pedestrian surveillance problem \cite{Toy1-square,Toy2-square},
where several groups of pedestrians or sportsmen walking together
in an interested surveillance area, positions of a group of pedestrians
are also highly correlated. In this case, the potential functions
$\psi\left(\boldsymbol{U}_{u}^{t},\boldsymbol{U}_{j}^{t}\right)$
in \eqref{eq:MRF} is given by $l_{2}$-norm priors in a exponential
function, 
\begin{equation}
\varphi(\boldsymbol{U}_{u}^{t},\boldsymbol{U}_{j}^{t})\propto\exp(-\frac{(d_{uj}^{t})^{2}}{2\sigma_{uj}^{2}}),\label{eq:Pedestrian}
\end{equation}
where $\sigma_{uj}^{2}$ is variance of inter-user distance.

\subsubsection{Communities with Few Distant Users}

For most of multi-users tracking and localization problem, the interaction
potential functions can be modeled in a quadratic prior as in \eqref{eq:Pedestrian},
however, the $l_{1}$-norm priors such as $d_{uj}^{t}$ also can be
used to limit the penalty computed on few of users far from the communities
users, which is used in large sensor network \cite{Toy1-norm,Toy2-norm},
i.e., 
\begin{equation}
\varphi\left(\boldsymbol{U}_{u}^{t},\boldsymbol{U}_{j}^{t}\right)\propto\exp(-\frac{d_{uj}^{t}}{2\sigma_{uj}^{2}}).\label{eq:Pedestrian-1}
\end{equation}

\section{Bayesian Multi-user Direct Localization and Tracking}

In this section, we incorporate the ST-MRF model and design an MUDLT
algorithm to achieve direct estimation of the positions of mobile
multiple users with the assistance of multi-RISs. After this, we analyze
the ST-PEB in multi-user dynamic localization system, which is leveraged
to optimize the passive BF at multi-RISs to provide optimal positioning
services. To better showcase our proposed algorithm, we assume a specialized
ST-MRF in this section, i.e., the spatial correlation of user positions
is only manifested among adjacent users. In this case, the MRF in
(xx) is given by
\begin{align*}
p(\boldsymbol{U}) & =p(\boldsymbol{U}^{1})\prod_{t=1}^{T}p(\boldsymbol{U}^{t}|\boldsymbol{U}^{t-1}),\\
 & =\prod_{u=1}^{U}p(\boldsymbol{U}_{u}^{1})p(\boldsymbol{U}_{u+1}^{1}|\boldsymbol{U}_{u}^{1})\\
 & \times\prod_{t=1}^{T}\prod_{u=1}^{U}p(\boldsymbol{U}_{u}^{t}|\boldsymbol{U}_{u}^{t-1})\prod_{i=\{1,-1\}}\psi(\boldsymbol{U}_{u}^{t},\boldsymbol{U}_{u+i}^{t}).
\end{align*}

\begin{figure}[tbh]
\centering{}\includegraphics[scale=0.25]{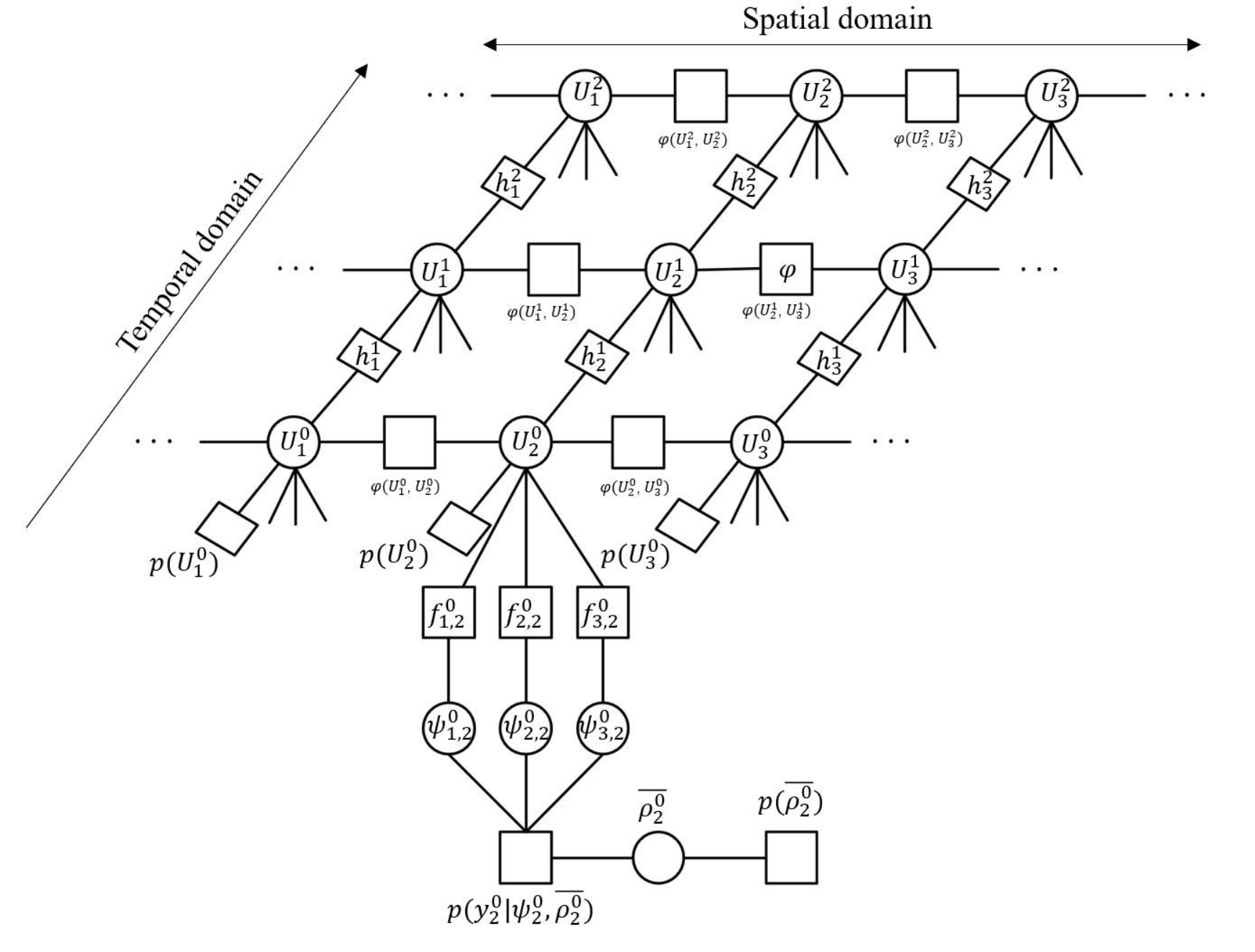}\caption{The message passing model of MRF \label{Fig. 2}}
\end{figure}

\subsection{MPDL Algorithm Framework}

It is a challenging task for direct position estimation in an RIS
system\cite{Bayesian9772371}. To address that, we propose a message-passing
mechanism for approximate computation. Specifically, the MPDL model
primarily consists of three modules as illustrated in Fig. 2:
\begin{itemize}
\item Module 1: AoA estimation , which performs AoA estimation from the
received signals.
\item Module 2: Message passing in temporal domain, where the transmitted
messages include both the previous temporal and spatial location information.
\item Module 3: Message passing in spatial domain, where the transmitted
messages include the current time and the previous time's location
information, and this spatial dimension message passing is bidirectional.
\end{itemize}
For these three modules, Module 1 employs a structure similar to the
MULT algorithm. Module 2 constructs a conditional probability model
using the MRF model in the temporal domain, enabling the transmission
of information regarding the previous temporal and spatial locations
through the introduction of this probability model. These two modules
are connected by factor node $p(\psi_{n,u}^{t}|\boldsymbol{U}_{u}^{t})$
and variable node $\psi_{n,u}^{t}$ according to the geometric relationship.
And Module 3 introduces a bidirectional message passing approximation
algorithm to achieve message passing in the spatial dimension. In
the following first subsection, the details of of Module 1 and Module
2 will be provided, while the Module 3 is illustrated in the second
subsection.

Some notations are described as follow. The $\boldsymbol{m}_{a\rightarrow b}$
and $\mathit{\boldsymbol{\mathit{\Sigma}}}_{a\rightarrow b}$ are
the mean vector and the covariance matrix of message message from
node $a$ to $b$. We denote the factor node $p(\psi_{n,u}^{t}|\boldsymbol{U}_{u}^{t})$
as $f_{n.u}^{t}$. And $P(\boldsymbol{U}_{u}^{t}|\boldsymbol{U}_{u}^{t-1})$
denote as $h_{u}^{t}$, where $E$ represents all users except the
$n$-th user in the MRF. Such that $\varphi(\boldsymbol{U}_{u}^{t},\boldsymbol{U}_{j}^{t})$
can be further simplified to $\varphi(\boldsymbol{U}_{u}^{t},\boldsymbol{U}_{u-1}^{t})$
or $\varphi(\boldsymbol{U}_{u}^{t},\boldsymbol{U}_{u+1}^{t})$, with
message passing based on the neighbor users. 

\subsubsection{Message passing in spatial domain for Module 1 and 3}

The AoA estimation module is designed to estimate users' AoAs ${\psi_{n,u}^{t}}$
which are further utilized for position tracking. With the prior probability
distribution of AoA can be designed as a Von Mises (VM) distribution,
then the AoA ${\psi_{n,u}^{t}}$ , equivalent channel gain $\bar{\rho}_{n,u}^{t}$
and their covariances can be estimated by VLASE \cite{VALSE}. Then
we get the message AoA module to position as $\boldsymbol{m}_{\psi_{n,u}^{t}\shortrightarrow f_{n.u}^{t}}(\psi_{n,u}^{t})$
and $\boldsymbol{\mathit{\Sigma}}_{\psi_{n,u}^{t}\shortrightarrow f_{n.u}^{t}}$.

Taking into account the spatial correlation among multiple users'
locations. After $\boldsymbol{m}_{\psi_{n,u}^{t}\shortrightarrow f_{n.u}^{t}}(\psi_{n,u}^{t})$
be calculated, the module 3 need to equires a re-computation in conjunction
with a spatio-temporal probability model. In spatial domain, the aim
is to analyze the correlations across multiple users. In spatial perspective,
we intend to examine the interrelationships among different users.
By considering these factors comprehensively, within the tracking
of each user, we simultaneously incorporate prior information from
the previous temporal step and associated users. This leads us to
introduce an advanced spatial correlation tracking algorithm. Building
upon this foundation, we have refined the corresponding message-passing
mechanism as follows, the massage passing as detailed below.

(1) Messages from $f_{n.u}^{t}$ to $\boldsymbol{U}_{u}^{t}$: For
$\forall t,1\leq u\leq U$, the message from $\{\psi_{n,u}^{t}\}$
to $\boldsymbol{U}_{u}^{t}$ is given by

\begin{equation}
\boldsymbol{m}_{f_{n.u}^{t}\shortrightarrow\boldsymbol{U}_{u}^{t}}(\boldsymbol{U}_{u}^{t})\propto\varint p(\psi_{n,u}^{t}|\boldsymbol{U}_{u}^{t})\boldsymbol{m}_{\psi_{n,u}^{t}\shortrightarrow f_{n.u}^{t}}(\psi_{n,u}^{t}),\label{eq:message1}
\end{equation}
where $\boldsymbol{m}_{\psi_{n,u}^{t}\shortrightarrow f_{n.u}^{t}}(\psi_{n,u}^{t})$
is the VM distribution

\begin{equation}
\boldsymbol{m}_{\psi_{n,u}^{t}\shortrightarrow f_{n.u}^{t}}(\psi_{n,u}^{t})=\frac{1}{2\pi I(\kappa)}exp(\kappa cos(\psi_{n,u}^{t}-\mu)),\label{eq:message2}
\end{equation}
here $\mu$ and $\kappa$ are the mean direction and concentration
parameters respectively.

(2) After receiving the message from the AoA module, the user will
further implement message passing in the spatial dimension as follows.
And messages along the Simplified MRF ${\boldsymbol{U}_{u}^{t}}$
is given by

\begin{equation}
\boldsymbol{m}_{\boldsymbol{U}_{u}^{t}\shortrightarrow\varphi\text{(}\boldsymbol{U}_{u}^{t},\boldsymbol{U}_{u+1}^{t})}(\boldsymbol{U}_{u}^{t})\propto\boldsymbol{m}_{\varphi\text{(}\boldsymbol{U}_{u}^{t},\boldsymbol{U}_{j}^{t})\shortrightarrow\boldsymbol{U}_{u}^{t}}(\boldsymbol{U}_{u}^{t})\mathcal{G}_{u}^{t}\text{(}\boldsymbol{U}_{u}^{t})p(\boldsymbol{U}_{u}^{t}),\label{eq:message3}
\end{equation}
here $j\in\{u+1,u-1\}$, and for $\forall t$, the message from $\psi_{n,u}^{t}$
to $\boldsymbol{U}_{u+1}^{t}$ is $\mathcal{G}_{u}^{t}\text{(}\boldsymbol{U}_{u}^{t})$
, and $\boldsymbol{m}_{h_{u}^{t}\shortrightarrow U_{u}^{t}}(\boldsymbol{U}_{u}^{t})$
is the message from factor $h_{u}^{t}$ to variable $\boldsymbol{U}_{u}^{t}$,
here $p(\boldsymbol{U}_{u}^{0})=\frac{1}{a-b},\boldsymbol{U}_{u}^{t}\in[a,b]$
is uniform distribution prior and $p(\boldsymbol{U}_{u}^{t})=\boldsymbol{m}_{h_{u}^{t-1}\shortrightarrow\boldsymbol{U}_{u}^{t-1}}(\boldsymbol{U}_{u}^{t}),t\geqq1$
, where $\boldsymbol{m}_{\varphi\text{(}\boldsymbol{U}_{u}^{t},\boldsymbol{U}_{u+1}^{t})\shortrightarrow\boldsymbol{U}_{u+1}^{t}}(\boldsymbol{U}_{u+1}^{t})$
is 

\begin{equation}
\begin{array}{cc}
\boldsymbol{m}_{\varphi\text{(}\boldsymbol{U}_{u}^{t},\boldsymbol{U}_{u+1}^{t})\shortrightarrow\boldsymbol{U}_{u+1}^{t}}(\boldsymbol{U}_{u+1}^{t})=\intop_{\boldsymbol{U}_{u}^{t}}\boldsymbol{m}_{\boldsymbol{U}_{u}^{t}\shortrightarrow\varphi\text{(}\boldsymbol{U}_{u}^{t},\boldsymbol{U}_{u+1}^{t})}(\boldsymbol{U}_{u}^{t})\\
\times\varphi\text{(}\boldsymbol{U}_{u}^{t},\boldsymbol{U}_{u+1}^{t}),
\end{array}\label{eq:message_forward-2}
\end{equation}
then it can be approached by Gaussian distribution 

\begin{equation}
\begin{array}{cc}
\boldsymbol{m}_{\varphi\text{(}\boldsymbol{U}_{u}^{t},\boldsymbol{U}_{u+1}^{t})\shortrightarrow\boldsymbol{U}_{u+1}^{t}}(\boldsymbol{U}_{u+1}^{t})\propto\\
\mathcal{N}(\boldsymbol{U}_{u+1}^{t};\boldsymbol{m}_{\varphi\text{(}\boldsymbol{U}_{u}^{t},\boldsymbol{U}_{u+1}^{t})\shortrightarrow\boldsymbol{U}_{u+1}^{t}},\boldsymbol{\mathit{\Sigma}}_{\varphi\text{(}\boldsymbol{U}_{u}^{t},\boldsymbol{U}_{u+1}^{t})\shortrightarrow\boldsymbol{U}_{u+1}^{t}}),
\end{array}\label{eq:message_forward-2-1}
\end{equation}
here and $\boldsymbol{m}_{\varphi\text{(}\boldsymbol{U}_{u}^{t},\boldsymbol{U}_{u+1}^{t})\shortrightarrow\boldsymbol{U}_{u+1}^{t}}$
and $\boldsymbol{\mathit{\Sigma}}_{\varphi\text{(}\boldsymbol{U}_{u}^{t},\boldsymbol{U}_{u+1}^{t})\shortrightarrow\boldsymbol{U}_{u+1}^{t}}$
are mean and variance. where $\mathcal{G}_{n}^{t}\text{(}\boldsymbol{U}_{u}^{t})=\prod_{j=1}^{K}\boldsymbol{m}_{f_{n,j}^{t}\shortrightarrow\boldsymbol{U}_{u}^{t}}(\boldsymbol{U}_{u}^{t})$
can be approximated using the law of the central limit theorem, so

\begin{equation}
\mathcal{G}_{u}^{t}\text{(}\boldsymbol{U}_{u}^{t})=\mathcal{N}(\boldsymbol{U}_{u}^{t};\boldsymbol{m}_{\mathcal{G}_{u}^{t}},\boldsymbol{\mathit{\Sigma}}_{\mathcal{G}_{u}^{t}}),
\end{equation}
and $\boldsymbol{m}_{\mathcal{G}_{u}^{t}}$ and $\boldsymbol{\mathit{\Sigma}}_{\mathcal{G}_{u}^{t}}$
can be calculated by the gradient descent method (GDM) and the Taylor
series expansion.

(3) In the spatial dimension, the messages received by the user factor
from the AoA module $\boldsymbol{U}_{u}^{t}$ to $\psi_{n,u}^{t}$
, the messages sent back from $\boldsymbol{U}_{u}^{t}$ to the AoA
module $f_{n.u}^{t}$ , and message $f_{n.u}^{t}$ to $\psi_{n,u}^{t}$
are same as \cite{Bayesian9772371}.

\subsubsection{Message passing in temporal domain Module 2}

during the forward propagation of spatial messages, user location
information can not only be passed to the next user but also, during
the backward propagation, the user can leverage the estimated values
obtained from the forward pass for further location accuracy enhancement.
Therefore, we propose a bidirectional approximation algorithm.

Assuming\textbf{ $P(\boldsymbol{U}_{u}^{t-1}|\boldsymbol{U}_{u}^{t})$
}as $h_{u}^{t}$ , the message from $h_{u}^{t}$ to $\boldsymbol{U}_{u}^{t}$
is

\begin{equation}
\boldsymbol{m}_{h_{u}^{t}\shortrightarrow U_{u}^{t}}(\boldsymbol{U}_{u}^{t})\propto\int_{\boldsymbol{U}_{u}^{t-1}}\boldsymbol{m}_{\boldsymbol{U}_{u}^{t-1}\shortrightarrow h_{u}^{t}}(\boldsymbol{U}_{u}^{t-1})h_{u}^{t},
\end{equation}
we obtain

\begin{equation}
\begin{array}{cc}
\boldsymbol{m}_{\boldsymbol{U}_{u}^{t-1}\shortrightarrow h_{u}^{t}}(\boldsymbol{U}_{u}^{t-1})\propto\boldsymbol{m}_{\varphi\text{(}\boldsymbol{U}_{j}^{t},\boldsymbol{U}_{u}^{t})\shortrightarrow\boldsymbol{U}_{u}^{t}}(\boldsymbol{U}_{u}^{t})\times\\
p(\boldsymbol{U}_{u}^{t-1})\mathcal{G}_{u}^{t}\text{(}\boldsymbol{U}_{u}^{t-1}),
\end{array}\label{eq:message_forward}
\end{equation}
here $j\in\{u+1,u-1\}$, the ${\boldsymbol{U}_{u}^{t}}$ in time is
given by

\begin{equation}
\begin{array}{cc}
\boldsymbol{m}_{h_{u}^{t}\shortrightarrow\boldsymbol{U}_{u}^{t}}(\boldsymbol{U}_{u}^{t})\propto\int_{\boldsymbol{U}_{u}^{t-1}}\boldsymbol{m}_{\varphi\text{(}\boldsymbol{U}_{j}^{t-1},\boldsymbol{U}_{u}^{t-1})\shortrightarrow\boldsymbol{U}_{u}^{t-1}}(\boldsymbol{U}_{u}^{t-1})\\
\times p(\boldsymbol{U}_{u}^{t})\mathcal{G}_{u}^{t}\text{(}\boldsymbol{U}_{u}^{t}),
\end{array}\label{eq:message_forward-1}
\end{equation}
which can be approached by Gaussian distribution

\[
\boldsymbol{m}_{h_{u}^{t}\shortrightarrow\boldsymbol{U}_{u}^{t}}(\boldsymbol{U}_{u}^{t})\propto\mathcal{N}(\boldsymbol{U}_{u}^{t};\boldsymbol{m}_{h_{u}^{t}\shortrightarrow\boldsymbol{U}_{u}^{t}},\boldsymbol{\mathit{\Sigma}}_{h_{u}^{t}\shortrightarrow\boldsymbol{U}_{u}^{t}}),
\]
and $\boldsymbol{m}_{h_{u}^{t}\shortrightarrow\boldsymbol{U}_{u}^{t}}$and
$\boldsymbol{\mathit{\Sigma}}_{h_{u}^{t}\shortrightarrow\boldsymbol{U}_{u}^{t}}$
are mean and variance.

\subsection{BCRB Optimization for Predictive PBF Design }

\begin{algorithm}[tbh]
\caption{MUDLT Algorithm\label{Algorithm MUDLT}}

\textbf{Input:} Observed signal $\mathbf{y}_{u}^{1},...,\mathbf{y}_{u}^{t},\forall u$,
the positions of RISs and BS ${\ensuremath{\boldsymbol{U}_{1},...\boldsymbol{U}_{N}}}$
and $\boldsymbol{U}_{B}$ Initialize users position $\boldsymbol{m}_{h_{u+1}^{0}\shortrightarrow\boldsymbol{U}_{u}^{0}}(\boldsymbol{U}_{u}^{0})$
and with its covariance matrix $\boldsymbol{\mathit{\Sigma}}_{h_{u+1}^{t}\shortrightarrow\boldsymbol{U}_{u}^{t}},\forall u$.

\textbf{Output: }Users position estimation $\boldsymbol{m}_{h_{u+1}^{t}\shortrightarrow\boldsymbol{U}_{u}^{t}}(\boldsymbol{U}_{n}^{t})$
at time $t$, the estimation of AoA and equivalent channel gain $\bar{\rho}_{n,u}^{t}$
and $\psi_{n,u}^{t}$ , $u\in[1,U]$ and $n\in[1,N]$.
\begin{enumerate}
\item \textbf{for} $t=1$ to $T$ do
\item $\text{\ensuremath{\quad}}$\textbf{for} $u=1$ to $U$ do
\item $\ensuremath{\quad}\quad$ Calculate $\boldsymbol{m}_{\psi_{n,u}^{t}\shortrightarrow f_{n.u}^{t}}(\psi_{n,u}^{t})$
and $\boldsymbol{m}_{f_{n.u}^{t}\shortrightarrow\boldsymbol{U}_{u}^{t}}(\boldsymbol{U}_{u}^{t})$
\item $\ensuremath{\quad}\quad$by \eqref{eq:message2}.
\item $\quad\quad$ \textbf{while not converge do}
\item $\quad\quad\quad$\textbf{\%Module} \textbf{1 and module 2 with fixed
PBF}
\item $\quad\quad\quad\quad$For $\forall u,n$, update $\boldsymbol{m}_{\boldsymbol{U}_{u}^{t}\shortrightarrow f_{n.u}^{t}}(\psi_{n,u}^{t})$
and
\item $\quad\quad\quad\quad$$\boldsymbol{m}_{f_{n.u}^{t}\shortrightarrow\psi_{n,u}^{t}}(\psi_{n,u}^{t})$
with their covariance
\item $\quad\quad\quad\quad$matrix similar with \cite{Bayesian9772371}.
\item $\quad\quad\quad\quad$Calculate $\forall u,n$, update $\psi_{n,u}^{t}$
and $\bar{\rho}_{n,u}^{t}$ by
\item $\quad\quad\quad\quad$VALSRE.
\item $\quad\quad\quad\quad$For $\forall u,n$, update $\boldsymbol{m}_{\psi_{n,u}^{t}\shortrightarrow f_{n.u}^{t}}(\psi_{n,u}^{t})=\psi_{n,u}^{t}$
.
\item $\quad\quad\quad\quad$For $\forall u,n$, update $\boldsymbol{m}_{f_{n.u}^{t}\shortrightarrow\boldsymbol{U}_{u}^{t}}(\psi_{n,u}^{t})=\psi_{n,u}^{t}$
by
\item $\quad\quad\quad\quad$\eqref{eq:message1} .
\item $\quad\quad\quad$\textbf{\%Module} \textbf{3 with fixed PBF}
\item $\quad\quad\quad\quad$For $\forall u,n$, update $\boldsymbol{m}_{\varphi\text{(}\boldsymbol{U}_{u}^{t},\boldsymbol{U}_{u+1}^{t})\shortrightarrow\boldsymbol{U}_{u+1}^{t}}(\boldsymbol{U}_{u+1}^{t})$
by \eqref{eq:message_forward-2}.
\item $\quad\quad\quad\quad$For $\forall u,n$, update $\boldsymbol{m}_{\boldsymbol{U}_{u}^{t-1}\shortrightarrow h_{u}^{t}}(\boldsymbol{U}_{u}^{t-1})$
by \eqref{eq:message_forward}.
\item $\quad\quad$\textbf{end while}
\item $\quad\quad$\textbf{\%PBF Optimization with estimated users}
\item $\quad\quad\quad$\textbf{position}
\item $\quad\quad\quad$For $\forall u,n$ approach calculate $\boldsymbol{m}_{\boldsymbol{U}_{u}^{t}\shortrightarrow h_{u}^{t+1}}(\boldsymbol{U}_{u}^{t})$
\item $\quad\quad\quad$by \eqref{eq:message_forward} in time $t$.
\item $\quad\quad\quad$Solve problem \eqref{eq:Prob-1} and obtain the
optimal
\item $\quad\quad\quad$PBF .
\item $\text{\ensuremath{\quad}}$\textbf{end for}
\item \textbf{end for}
\end{enumerate}
\end{algorithm}

As depicted in field of user prediction researches \cite{CRB}, a
rough predictive PBF, like received SNR maximization criterion, will
lead to tracking performance degradation or even tracking failure.
Two main challenges lies in the works of predictive PBF design. First
is that the predictive PBF design based on the optimal design metric
such as MSE or maximization of likelihood function, is extremely difficult
and computation-cost to solve a PBF solution. Another challenge is
the predication of users position in the future time is hard to obtained.
Most of existing works adopt the current position for predictive PBF
design approximately, i.e., $\boldsymbol{m}(\boldsymbol{U}_{u}^{t+1})\approx\boldsymbol{m}(\boldsymbol{U}_{u}^{t})$,
where $\boldsymbol{m}(\boldsymbol{U}_{u}^{t})$ is an abbreviation
of user position in \eqref{eq:tr_ue_pos}. This method will provoke
the tracking accuracy loss especially for the less snapshots scenario.

In this paper, we employ another solvable optimal design metric in
terms of BCRB for RISs predictive PBF design, which provides a lower
bound of MSE of users position. Combining with the accurate users
predictive position obtained in the proposed MUDLT algorithm in \eqref{Algorithm MUDLT},
our predictive RISs PBF design optimization problem is given by

\begin{align}
 & \min_{\boldsymbol{\lambda}^{t+1}}\:BCR(\boldsymbol{\lambda}^{t+1},\boldsymbol{m}(\boldsymbol{U}_{u}^{t+1}))\label{eq:Prob-1}\\
 & s.t.\quad\left|\boldsymbol{\lambda}_{n,i}^{t}\right|=1,i\in[1,M_{R}],\nonumber 
\end{align}
where $\boldsymbol{\lambda}^{t+1}$ is the PBF tandem vector for $R$
RISs. The objective function the trace of inverse of equivalent Fisher
information matrix of users predictive positions, i.e., $BCR(\boldsymbol{\lambda}^{t+1},\boldsymbol{m}(\boldsymbol{U}_{u}^{t+1}))=\textrm{tr}\left(\mathbf{J}_{e}^{-1}(\boldsymbol{\lambda}^{t+1},\boldsymbol{m}(\boldsymbol{U}_{u}^{t+1}))\right)$.
Define the unknown parameter $\boldsymbol{\eta}=[\psi^{t},\boldsymbol{\bar{\rho}}^{t}]^{T}$
and noiseless signal $\boldsymbol{\mu}^{t}=\left[\mathbf{y}_{1}^{t}-\mathbf{n}_{1}^{t},\ldots,\mathbf{y}_{U}^{t}-\mathbf{n}_{U}^{t}\right]^{T}$
, the FIM is given by

\begin{equation}
\mathbf{J}_{e}(\boldsymbol{\lambda}^{t+1},\boldsymbol{m}(\boldsymbol{U}_{u}^{t+1}))=\frac{P}{\sigma^{2}}\mathbf{T}^{T}\mathbf{J}_{\boldsymbol{\eta}}(\boldsymbol{\lambda}^{t+1},\boldsymbol{m}(\boldsymbol{U}_{u}^{t+1}))\mathbf{T}+\mathbf{J}_{\boldsymbol{p}},\label{eq:J_e}
\end{equation}
where $\frac{P}{\sigma^{2}}$ is SNR of received signal, $\mathbf{T}$
represents transformation matrix with $\mathbf{T}=\frac{\partial\mathbf{\psi^{t}}}{\partial\boldsymbol{m}(\boldsymbol{U}_{u}^{t+1})}$,
$\mathbf{J}_{\boldsymbol{p}}$ is EFIM of spatiotemporal prior and
$\mathbf{J}_{\boldsymbol{\eta}}$ is EFIM of $\boldsymbol{\eta}$,
which is given by

\begin{equation}
\mathbf{J}_{\boldsymbol{\eta}}(\boldsymbol{\lambda}^{t+1},\boldsymbol{m}(\boldsymbol{U}_{u}^{t+1}))=\mathfrak{R}\left\{ \frac{\partial(\boldsymbol{\mu}^{t}){}^{H}}{\partial\boldsymbol{\eta}}\frac{\partial(\boldsymbol{\mu}^{t})}{\partial\boldsymbol{\eta}}\right\} ,\label{eq:J_eta}
\end{equation}

\begin{equation}
\mathbf{J}_{\boldsymbol{p}}=-\mathbb{E}\left\{ \frac{\partial(\log(p(\boldsymbol{U}_{u}^{t}))){}^{H}}{\partial\boldsymbol{m}(\boldsymbol{U}_{u}^{t+1})}\frac{\partial(\log(p(\boldsymbol{U}_{u}^{t})))}{\partial\boldsymbol{m}(\boldsymbol{U}_{u}^{t+1})}\right\} ,
\end{equation}
and, we have

\begin{equation}
\begin{array}{cc}
\frac{\partial\mathfrak{R}\{\boldsymbol{\mu}^{t}\}}{\partial\psi_{n,u}^{t}}=\sum_{n=1}^{N}-|\bar{\rho}_{n,u}^{t}|\pi(n-1)\sin(\pi(n-1)\psi_{n,u}^{t}\\
+\arg(\bar{\rho}_{n,u}^{t}),
\end{array}\label{eq:message3-1}
\end{equation}

\begin{equation}
\frac{\partial\mathfrak{R}\{\boldsymbol{\mu}^{t}\}}{\partial|\bar{\rho}_{n,u}^{t}|}=\cos(\pi(n-1)\psi_{n,u}^{t}+\arg(\bar{\rho}_{n,u}^{t}),
\end{equation}

\begin{equation}
\frac{\partial\mathfrak{R}\{\boldsymbol{\mu}^{t}\}}{\partial\arg(\bar{\rho}_{n,u}^{t})}=-|\bar{\rho}_{n,u}^{t}|\sin(\pi(n-1)\psi_{n,u}^{t}+\arg(\bar{\rho}_{n,u}^{t}).
\end{equation}
The problem in \eqref{eq:J_e} is highly nonconvex due to its nonconvex
objective function with unit modulus constraint, which is hard to
solve in general. To address this problem, we first employ the matrix
lifting technique to transform the objective function into a convex
form. Then the unit modulus constraint can be handled by SDR method
\cite{SDR}. Denote PBF matrix as $\mathrm{\Pi}_{t+1}=\boldsymbol{\lambda}_{n}^{t+1}\left(\boldsymbol{\lambda}_{n}^{t+1}\right)^{H}$,
$\mathbf{J}_{\boldsymbol{\eta}}$ in \eqref{eq:J_eta} can be written
as

\begin{equation}
\mathbf{J}_{\boldsymbol{\eta}}(\mathrm{\Pi}_{t+1},\boldsymbol{m}(\boldsymbol{U}_{u}^{t+1}))=\alpha_{u}\xi_{n_{b}}^{2}\xi_{n_{u}}^{2}\boldsymbol{A}_{R}^{H}\mathrm{\Pi}_{t+1}\boldsymbol{A}_{R},
\end{equation}
and

\begin{equation}
\alpha_{u}=(\pi(n-1)\sin(\pi(n-1)\psi_{n,u}^{t})^{2},
\end{equation}

\begin{equation}
\boldsymbol{A}_{R}=\mathbf{a}_{R}^{H}(\psi_{RX,n}^{t},\psi_{RY,n}^{t})\mathbf{a}_{R}(\varphi_{RX,n},\varphi_{RY,n}).
\end{equation}
It is noticed that there is no RISs PBF matrix in prior FIM $\mathbf{J}_{\boldsymbol{p}}$
and $\mathbf{J}_{\boldsymbol{\eta}}(\mathrm{\Pi}_{t+1},\boldsymbol{m}(\boldsymbol{U}_{u}^{t+1}))$
is linear to PBF matrix $\mathbf{\mathrm{\Pi}}_{t+1}$, which lead
to a convex objective function. and. Taking into account SDR method
by dropping out the unit modulus constraint, the problem in \eqref{eq:Prob-1}
can be reformulated by

\begin{equation}
\min_{\mathrm{\Pi}_{t+1}}\:BCRB(\mathrm{\Pi}_{t+1},\boldsymbol{m}(\boldsymbol{U}_{u}^{t+1}))\;\;\;s.t.\:\mathrm{\Pi}_{t+1}\succeq\boldsymbol{0},\label{eq:BF opt}
\end{equation}
which can be efficiently solved by the CVX toolbox with an affordable
complexity of $\mathcal{O}\left(\left(URM_{R}\right)^{3.5}\right)$.

By executing MPDL algorithm and BCRB optimization with fixed positions
or PBF alternatively , the multi-user tracking and localization procedures
can be summarized in Algorithm MUDLT.

\section{NUMERICAL EXPERIMENTS}

We evaluate the multi-user tracking performance of proposed algorithm
in the multi-RIS-assisted system for the application of pedestrian
surveillance. The parameters are set up as follows: $N_{B}=32$, $N_{U}=16$
and $M_{R}=32$ with $N_{x}=4,N_{y}=8$, the unit direction vector
of the user antennas is $\boldsymbol{e}_{U}=[0,1,0]$. The pathloss
are obtained by $\xi_{n}=\frac{\lambda e^{-j\frac{2\pi d_{B,n}}{\lambda}}}{4\pi d_{B,n}}$
and $\text{\textgreek{\char6}}_{n,u}^{t}=\frac{\lambda e^{-j\frac{2\pi d_{n,u}}{\lambda}}}{4\pi d_{B,u}}$
with $\lambda=28$ GHz, $d_{B,n}$ and $d_{n,u}$ are the distances
from the BS to the $n$-th RIS and the $n$-th RIS to the $u$-user
respectively, the the AGWN noise is $\sigma_{n}^{2}=-120$ dBm, where
$\rho_{n}^{t}$ follows the model in \cite{tang2020wireless}. We
set the temporal and spatial correlation variances are $\boldsymbol{C}=\textrm{diag}([0.1,0.1,0.1]^{T})$
and $\boldsymbol{\mathit{\Sigma}}_{\varphi(\boldsymbol{U}_{u}^{t},\boldsymbol{U}_{j}^{t})}=\textrm{diag}([0.2,0.2,0.2]^{T})$
in \eqref{eq:time_pass} and \eqref{eq:Pedestrian}, while the temporal
prior \textbf{$\phi(\boldsymbol{U}_{u}^{1})$} follows uniform distribution
\textbf{$\phi(\boldsymbol{U}_{u}^{1})\sim\frac{1}{a-b},\boldsymbol{U}_{u}^{1}\in[a,b]$}
with $a=\boldsymbol{U}_{u}^{1}-0.5$ m and $b=\boldsymbol{U}_{u}^{1}+0.5$
m are the lower and uper bound. $R=6$ RISs and one BS located at$\left[20,-20,0\right]^{T}$,
$\left[20,20,0\right]^{T}$, $\left[15,20,0\right]^{T}$, $\left[15,-20,0\right]^{T}$,
$\left[25,20,0\right]^{T}$, $\left[25,-20,0\right]^{T}$ and $\boldsymbol{U}_{B}=\left[40,0,0\right]^{T}$
respectively, all in meters. Assume $K=3$ users roaming in a rectangular
area with speed of $1.5$ $m/s$ \cite{sun2016indoor} where x-axis
from$-10$ m to $10$ m and y-axis from $-10$ m to $10$ m. All simulation
results are demonstrated by root mean square error (RMSE) in meter
over 500 Monte-Carlo tests.

\subsection{Tracking Performance for Different Parameters }

\begin{figure}[tbh]
\begin{centering}
\includegraphics[scale=0.16]{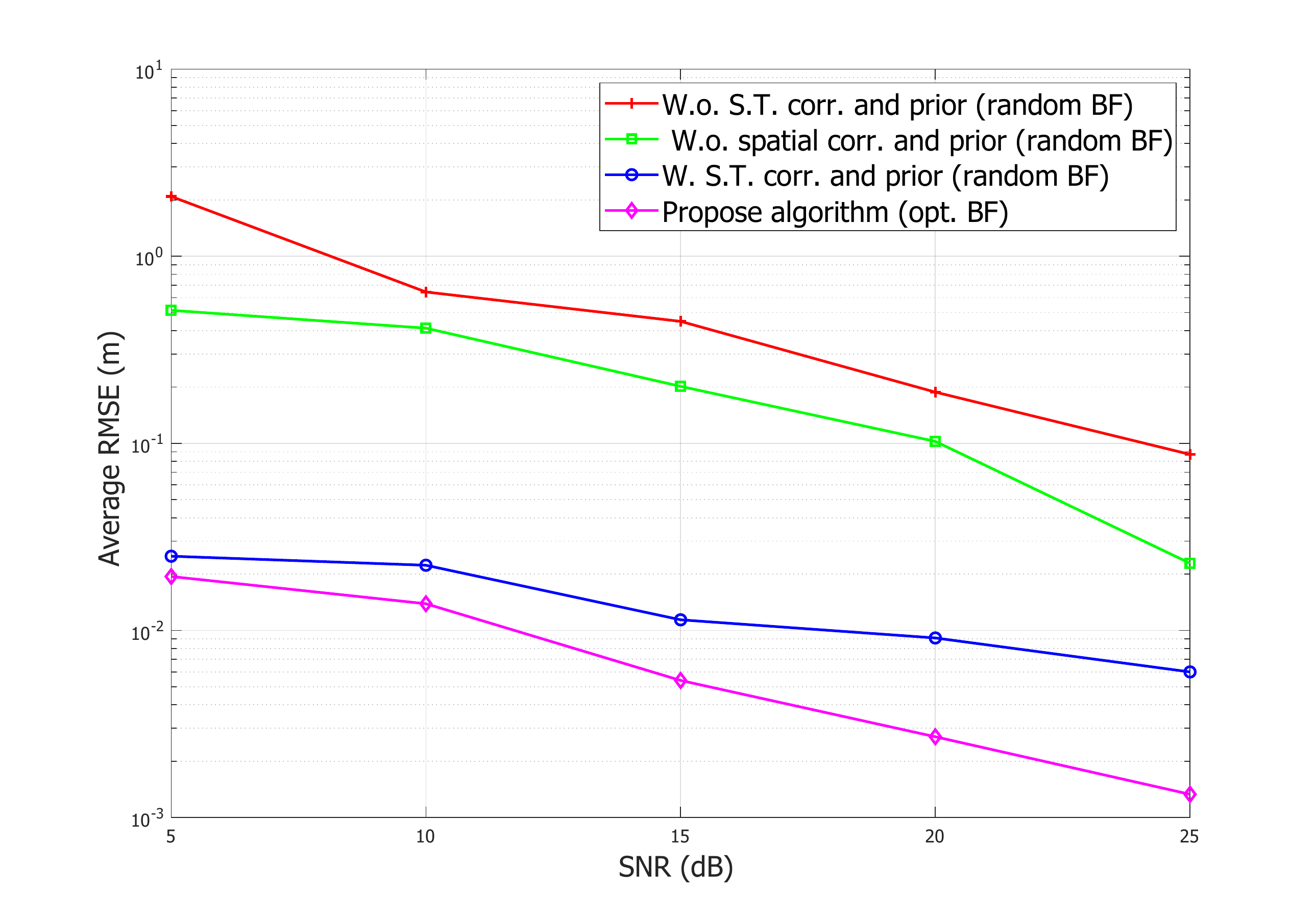}
\par\end{centering}
\caption{Averaged RMSE versus SNR for different schemes. \label{Fig. 3}}
\end{figure}
\begin{figure}[tbh]
\begin{centering}
\includegraphics[scale=0.16]{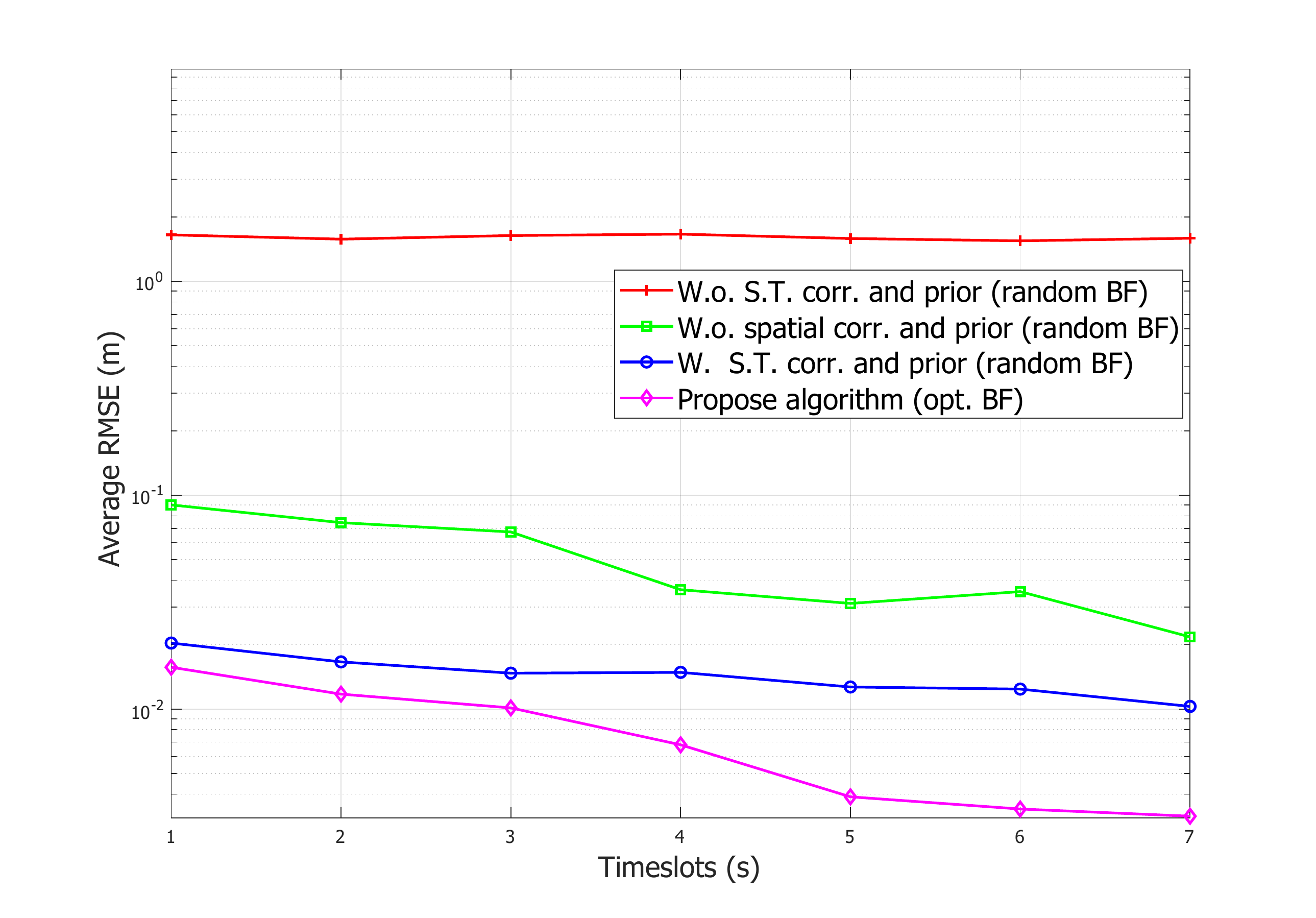}
\par\end{centering}
\caption{RMSE versus time slots for different schemes.\label{Fig. 4}}
\end{figure}

In Fig. \ref{Fig. 3}, we show multi-users' averaged RMSE versus SNR
for three baselines. Baseline 1: multi-user tracking only by measurements
without spatiotemporal correlation and prior (W.o. S.T. corr. and
prior); Baseline 2: similar to reference \cite{Bayesian9772371},
tracking with temporal correlation (W.o. spatial corr. and prior);
Baseline 3: tracking both with spatiotemporal correlation and prior
(W. S.T. and prior). It is noticed that all baselines use random BF
to highlight the significance of proposed BF prediction in \eqref{eq:BF opt}.
It shows that the proposed algorithm achieves the minimum RMSE compared
to various baselines. Moreover, the proposed algorithm can obtain
more performance gain over SNR compared to Baseline 3 due to the high
accuracy BF prediction. In Fig. \ref{Fig. 3}, we evaluate the averaged
RMSE for all users as the time goes by. Since the baseline 1 only
uses measurements from signal, the multi-user tracking performance
keeps in a same level at different time slots. While the proposed
algorithm outperforms the baselines again and obtains more performance
gain as time slots increasing. Therefore, it is paramount to design
the predictive BF and estimation algorithm jointly when multiple users
exhibit spatiotemporal correlations. 

\subsection{Tracking Performance for Practical Trajectory}

\begin{figure}[tbh]
\begin{centering}
\includegraphics[scale=0.16]{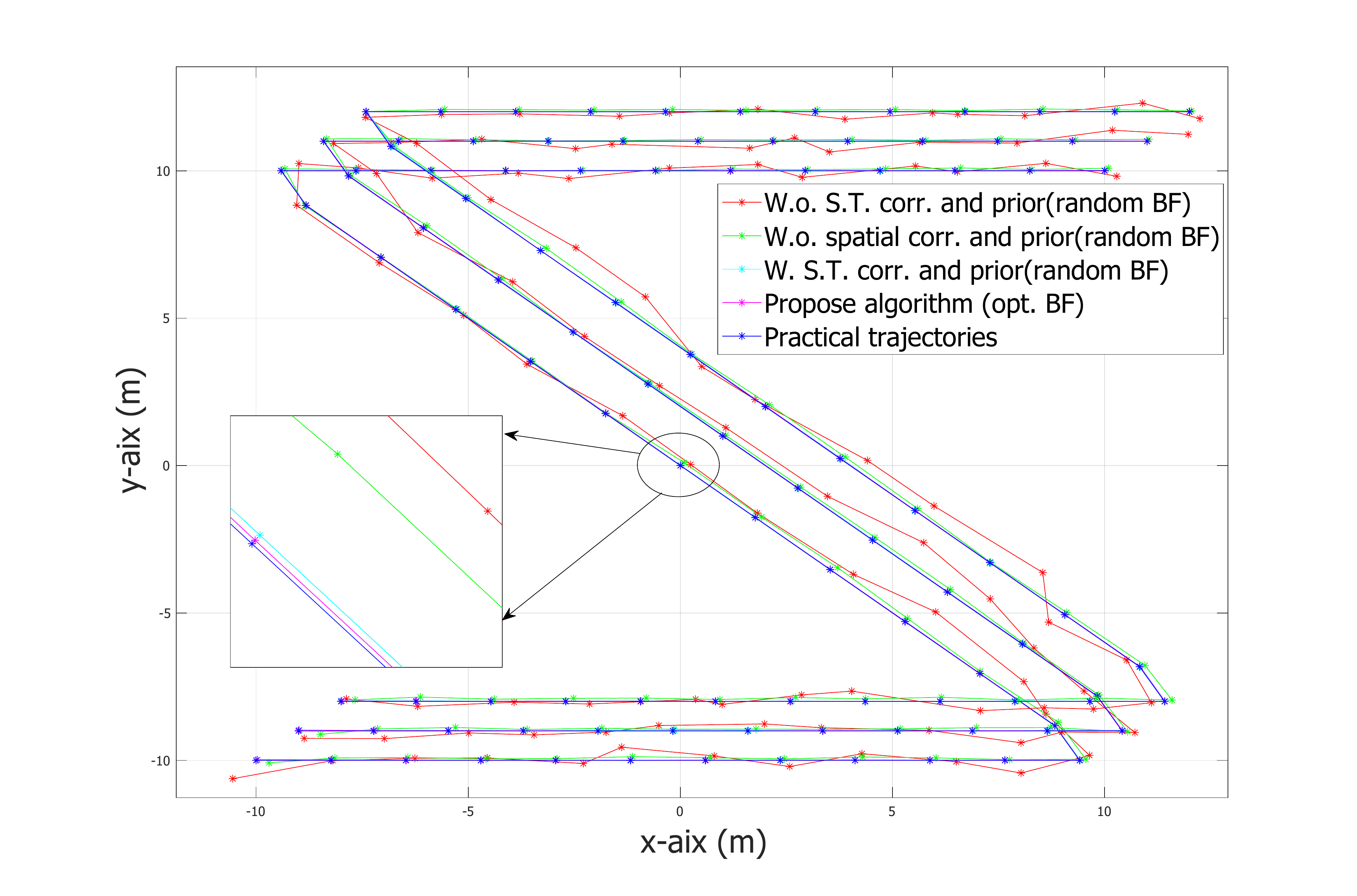}
\par\end{centering}
\caption{The practical trajectories and tracking performance of three users\label{Fig. 5}}
\end{figure}

Here we assume three users are both moving to a destination by a ``Z''
shaped trajectory. For example, the user 1 starts from $\left[-10,-10,0\right]^{T}$
m and end at $\left[10,10,0\right]^{T}$ m at the speed of $1.5$
m/s. In Fig. \ref{Fig. 5}, it is shown that the proposed algorithm
outperforms the baselines over the time. Besides that, we can find
that Baseline 1 and 2 have significant gap compared to Baseline 3
and proposed algorithm. This result implies that fully explore the
correlation among spatial and temporal domain can improve the tracking
performance remarkably.

\section{CONCLUSION}

In this paper, we investigate the problem of RIS-assisted localization
and tracking in multi-user systems. To explore the spatiotemporal
correlation among multiple users, we first adopt the ST-MRF model
to capture correlation and propose a novel MUDLT framework to obtain
high accuracy tracking performance. Specifically, in such framework,
the users position is first jointly estimated based on a message passing
method and then the RISs configuration is optimized from BCRB minimization
for futher superior performance. Simulation results demonstrate the
superiority of our proposed algorithm.

\appendices{}

\bibliographystyle{IEEEtran}
\bibliography{ref}

\end{document}